\documentclass[%
 reprint,
superscriptaddress,
bibnotes,
 amsmath,amssymb,
 aip,
]{revtex4-1}

\usepackage{graphicx}% Include figure files
\usepackage{dcolumn}% Align table columns on decimal point
\usepackage{bm}% bold math

\usepackage{hyperref}
\usepackage{ bbold }
\usepackage{natmove}
\usepackage{color}
\renewcommand{\vec}[1]{\mathbf{#1}}

\begin{document}
% \beginsupplement
\title{Inferring equilibrium transition rates from nonequilibrium protocols}

\preprint{APS/123-QED}
\author{Benjamin Kuznets-Speck}

\affiliation{Biophysics Graduate Group, University of California, Berkeley, CA, 94720, USA \looseness=-1}
\author{David T. Limmer}%
\email{dlimmer@berkeley.edu}
\affiliation{Chemistry Department, University of California, Berkeley, CA, 94720, USA \looseness=-1}
\affiliation{Chemical Sciences Division, Lawrence Berkeley National Laboratory, Berkeley, CA, 94720, USA \looseness=-1}
\affiliation{Material Sciences Division, Lawrence Berkeley National Laboratory, Berkeley, CA, 94720, USA \looseness=-1}
\affiliation{Kavli Energy NanoSciences Institute, University of California, Berkeley, CA, 94720, USA \looseness=-1}

\date{\today}

\begin{abstract}
We develop a theory for inferring equilibrium transition rates from trajectories driven by a time dependent force using results from stochastic thermodynamics. Applying the Kawasaki relation to approximate the nonequilibrium distribution function in terms of the equilibrium distribution function and the excess dissipation, we formulate a nonequilibrium transition state theory to estimate the rate enhancement over the equilibrium rate due to the nonequilibrium protocol. We demonstrate the utility of our theory in examples of pulling of harmonically trapped particles in 1 and 2 dimensions, as well as a semi-flexible polymer with a reactive linker in 3 dimensions. In all cases we find that we are able to infer the transition rates more effectively than phenomenological approaches based on Bell's law. We expect our thermodynamic approach will find use in both molecular simulation and force spectroscopy experiments. 
\end{abstract}

\maketitle

%  Click the title above to edit the author information and abstract

\thispagestyle{empty}
% \onecolumngrid
\appendix

%% NEED CITATIONS THROUGHOUT

Extracting thermodynamic information from molecular systems through irreversible and dissipative processes was made possible with the revelation of Jarzynski's equality. \cite{jarzynski1997nonequilibrium,crooks1999entropy, hummer2001free,hummer2005free}
However, inference of kinetic information, such as the intrinsic rate of molecular transitions, has remained more elusive.\cite{hughes2016physics} Although useful ways of extracting transition rates from single molecule force data exist, for example, they often rely on fitting to phenomenological expressions\cite{bell1978models} or specifying a simple model of the underlying conformational landscape. \cite{hummer2003kinetics, dudko2008theory}
Such theories also typically hinge on the limiting assumption that the driving forces are adiabatic, so that the molecule is assumed to be in instantaneous equilibrium with the experimental forces imposed on it. Here, we report that a molecule's bare equilibrium transition rate can be inferred from the statistics of the excess heat released during a nonequilibrium protocol. This result derives from expressions from stochastic thermodynamics\cite{seifert2012stochastic} and an extension of transition state theory into nonequilibrium regimes.

One of the most common methods for extracting rate information from nonequilibrium experiments and simulations employs the so-called Bell's law.\cite{bell1978models} Bell's phenomenological rate law postulates that the speed of a molecular transition is accelerated with an applied external force by a factor that varies exponentially 
\begin{equation}
  k_\lambda \approx k_0 e^{\beta \lambda x^\dag} , \,
\end{equation}
where $k_\lambda$ is the rate in the presence of the external force $\lambda$,  $k_0$ is the equilibrium rate, $x^\dag$ is the distance along the forcing direction between the reactant state and a putative transition state, and $\beta$ is the inverse temperature times Boltzmann's constant. Evans and Richie showed that such a form emerges from Kramer's theory of transition rates when considering the special case of pulling a simple harmonic molecule through a transition state adiabatically.\cite{evans1997dynamic} Dudko, Hummer and Szabo \cite{dudko2006intrinsic} developed additional expressions for the rate amplification based on different potentials and the assumption of adiabaticity, and more recently, introduced a model-free method of estimating the force-dependent transition rate with statistics from the rupture force distribution.\cite{dudko2008theory}   
Using a nonequilibrium response relation for reaction rates,\cite{kuznets2021dissipation} we provide a general perspective on the origin of Bell's law, from which we can develop systematic corrections. We explore the subsequent generalizations in a number of molecular systems with increasing complexity, and study the utility of  model-independent rate estimates that depend only on the statistics of the dissipation, a quantity that is measurable experimentally. 

To start, we demonstrate how Bell's law can be understood as a consequence of two distinct approximations, a transition state theory approximation and a near equilibrium approximation. Within transition state theory, the rate of a transition between two metastable states is 
\begin{equation}
k_\lambda = \nu p_\lambda (x^\dag)
\end{equation}
where $\nu$ is the probability flux through $x^\dag$  and $p_\lambda (x^\dag)$ the probability to reach a transition state, or dividing surface in phase space between metastable states.\cite{peters2017reaction} This expression is an approximation to the rate because $\nu$ in principle depends on $x^\dag$, and errors associated with this approximation can be minimized with a judicious choice of dividing surface.\cite{chandler1978statistical} In equilibrium, the probability to reach a transition state is given by a Boltzmann distribution, $p_0 (x^\dag)\propto \exp[-\beta F(x^\dag)]$, where $ F(x^\dag)$ is the free energy to reach $x^\dag$, resulting in the expected Arrhenius temperature dependence. 

While transition state theory in still applicable to system away from equilibrium,\cite{bouchet2016generalisation, banik2000generalized,kuznets2021dissipation} the likelihood of reaching the transition state is not generally known, rendering it difficult to employ. For a system initially in equilibrium and acted upon by an external force, the nonequilibrium distribution is encoded in the Kawaski relation,\cite{yamada1967nonlinear, morriss1985isothermal,crooks2000path,lesnicki2021molecular}
\begin{equation}
p_0(\vec{x}) = p_\lambda (\vec{x}) \left \langle e^{-\beta Q} \right \rangle_{\lambda,\vec{x}}
\end{equation}
where $p_0(\vec{x})$ is the initial equilibrium probability of configuration $\mathbf{x}$, $p_\lambda(\vec{x})$ is the nonequilibrium probability distribution, and the brackets $\left \langle \dots \right \rangle_{\lambda,\vec{x}}$ denote an average under the driving force $\lambda$ conditioned on ending at $\vec{x}$. The excess heat dissipated to the environment, $Q$, due to the driving force is given by
\begin{equation}
\label{eq:heat}
    Q(t) = \int_0^t dt' \dot{\vec{x}}(t') \cdot \boldsymbol{\lambda}(\vec{x}, t'),
\end{equation} 
which for a single molecule pulling experiment could be inferred from the force-extension curve. 

If the barrier to transitioning is large, we expect that the dominant change in the rate under an additional force is due to modulation of $p(x^\dag)$, leaving the location and flux through the transition state unchanged. Under these assumptions, we employ the Kawaski relationship together with transition state theory to relate the rates in the presence and absence of $\lambda$, 
\begin{equation}
\label{eq:cumulant} 
    \frac{k_\lambda}{k_0}  \approx \langle e^{-\beta Q} \rangle_{\lambda,x^\dag}^{-1} \, ,
\end{equation}
to the statistics of the dissipation.  In this limit, the Kawasaki response relation connects the transition rate amplification and the distribution of excess heat dissipated to the bath. A similar rate enhancement relation has been derived from path ensemble techniques.\cite{kuznets2021dissipation}
Under an assumption of small applied force, the equilibrium rate can be estimated with a cumulant expansion for small values of the $\beta Q$ as  
\begin{equation}
\label{Qexpansion}
    \ln k_0  \approx \ln k_\lambda - \beta \langle Q  \rangle  + \frac{\beta^2}{2} \langle \delta Q^2 \rangle, 
\end{equation}
which is our first main result. For simplicity of notation we drop the explicit condition on the averages in the heat. It implies that measuring the rate of a rare event, as determined by a mean first passage time to $x^\dag$, and accompanying heat statistics in a driven system, we can infer the rate of a rare event in thermal equilibrium. This result can be considered as a nonlinear response theory for the rate,\cite{sivak2012near} in which frenetic contributions are neglected.\cite{gao2019nonlinear} It is similar to higher order corrections to Bell's law valid for constant applied forces,\cite{abkenar2017dissociation, konda2011chemical} generalized to nonequilibrium protocols. 

Near-equilibrium, $\beta \langle Q \rangle \ll 1$, and for slow driving forces, $\dot{\lambda}\approx 0$, the heat dissipated up the transition state $x^\dag$ is well approximated by
$\langle Q \rangle_{\lambda,x^\dag} \approx \lambda x^\dag$, 
where we have included only terms first order in $\lambda$ and integrated Eq.~\ref{eq:heat} by parts. Such an approximation is good when the transition remains activated, so the dynamics along the transition path relax much faster than the transition waiting time. Substituting this approximation for the average heat into Eq.~\ref{Qexpansion} and truncating at first order, we find Bell's law. This is our second main result.

\begin{figure}[h]
\centering
\includegraphics[width=8.5cm]{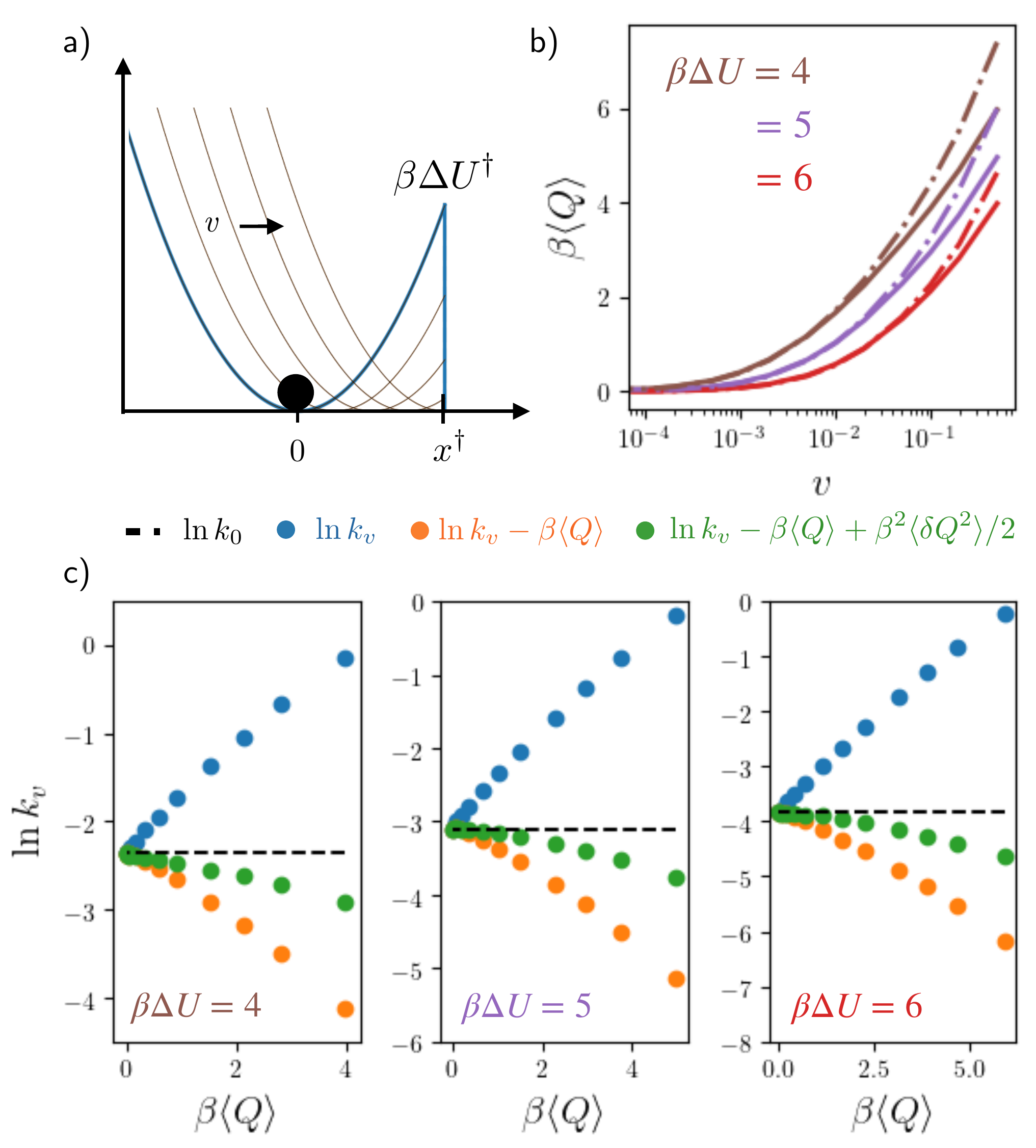}
\caption{Pulling on a harmonic molecule. (a) A cartoon of the time dependent potential, initial barrier height $\beta \Delta U^\dag$, and absorbing boundary condition, $x^\dag$.  (b) The heat dissipated gives Bell's law in the small loading rate, long rupture time limit. Dashed curves are Bell's law and the solid curves the full dissipation for three potential stiffnesses. (c) Estimates of the equilibrium escape rate for increasing barrier heights and a range of pulling velocities.
}
\label{fig1}
\end{figure}

Understanding the general thermodynamic origin of Bell's phenomenological law allows us to systematically improve upon it. In order to test such corrections, we have considered a hierarchy of models with increasing complexity. In each, we apply a simple force ramp, with constant velocity $v$ so that $\lambda \propto v$, and we measure the rate under this driving protocol, $k_v$, as a mean first passage time to an absorbing boundary condition. The first model we consider is a simple  overdamped particle trapped in a harmonic potential in one dimension, labeled $x$. The equation of motion is 
\begin{equation}
\dot{x} =\mu F(x) + \mu \lambda  + \sqrt{2 D} \eta
\end{equation}
where the mobility $\mu$ and diffusivity $D$ satisfy an Einstein relation $\beta D = \mu$,  $F$ is a conservative force, and $\eta$ is a Gaussian random variable with $\langle \eta \rangle=0$ and $\langle \eta(t) \eta(t') \rangle = \delta(t-t')$. The conservative force is $F = -a \kappa x$, and the particle is pulled with a linear ramp at loading rate $v$, such that $\lambda= a \kappa v t$  as depicted in Fig.~\ref{fig1}a. We fix $ \kappa = x^\dag = \beta  = \mu  = 1$ and vary $v=\{0,..,0.5 \}$ and $a=\{8,10,12\}$. Simulations are run from an initial condition at the origin and stopped upon crossing $x^\dag$ at $t = t_\mathrm{rup}$, where $k_v=1/\langle t_\mathrm{rup}\rangle$. Time is measured in units of $\tau=\kappa/\mu$. Associated with the location of the absorbing boundary condition is an increased potential energy, equal to $ \Delta U^\dag = a\kappa x^{\dag 2}/2$. We use a time step $t=10^{-2} \tau$ and average over $10^4$ trajectories. 

In Fig. ~\ref{fig1}b) we verify that the relationship between the heat and the argument of the exponential in Bell's law. Under a constant velocity force to lowest order in $v$, $Q \approx \kappa v t_\mathrm{rup}$, which is plotted against the full expression for $Q$. As expected, for small $v$, in which $\beta \langle Q \rangle \ll 1$, both estimates agree. Note that $\langle Q \rangle$ does not very linearly with $v$ because $t_\mathrm{rup}$ depends implicitly on loading rate. Figure ~\ref{fig1}c) illustrates the convergence of the equilibrium rate employing different estimators obtained from Eq.~\ref{Qexpansion}, for a range of pulling velocities and barrier heights. The Bell's law like rate estimate, correcting the rate with just the mean heat, approaches the true equilibrium rate $\ln k_0$ from below. Within the validity of transition state theory, this behavior can be understood as a consequence of Jensen's inequality, applied to Eq. ~\ref{eq:cumulant}. Including additional dissipation statistics as in the full expression in Eq. \ref{Qexpansion} yields a faster convergence to the equilibrium rate for all barrier heights considered. Significant deviations from Eq. \ref{Qexpansion} occur when the dissipation is comparable to the size of the energy barrier, in which case the barrier is degraded enough for the event to no longer be rare. 

\begin{figure}[b]
\centering
\includegraphics[width=8.5cm]{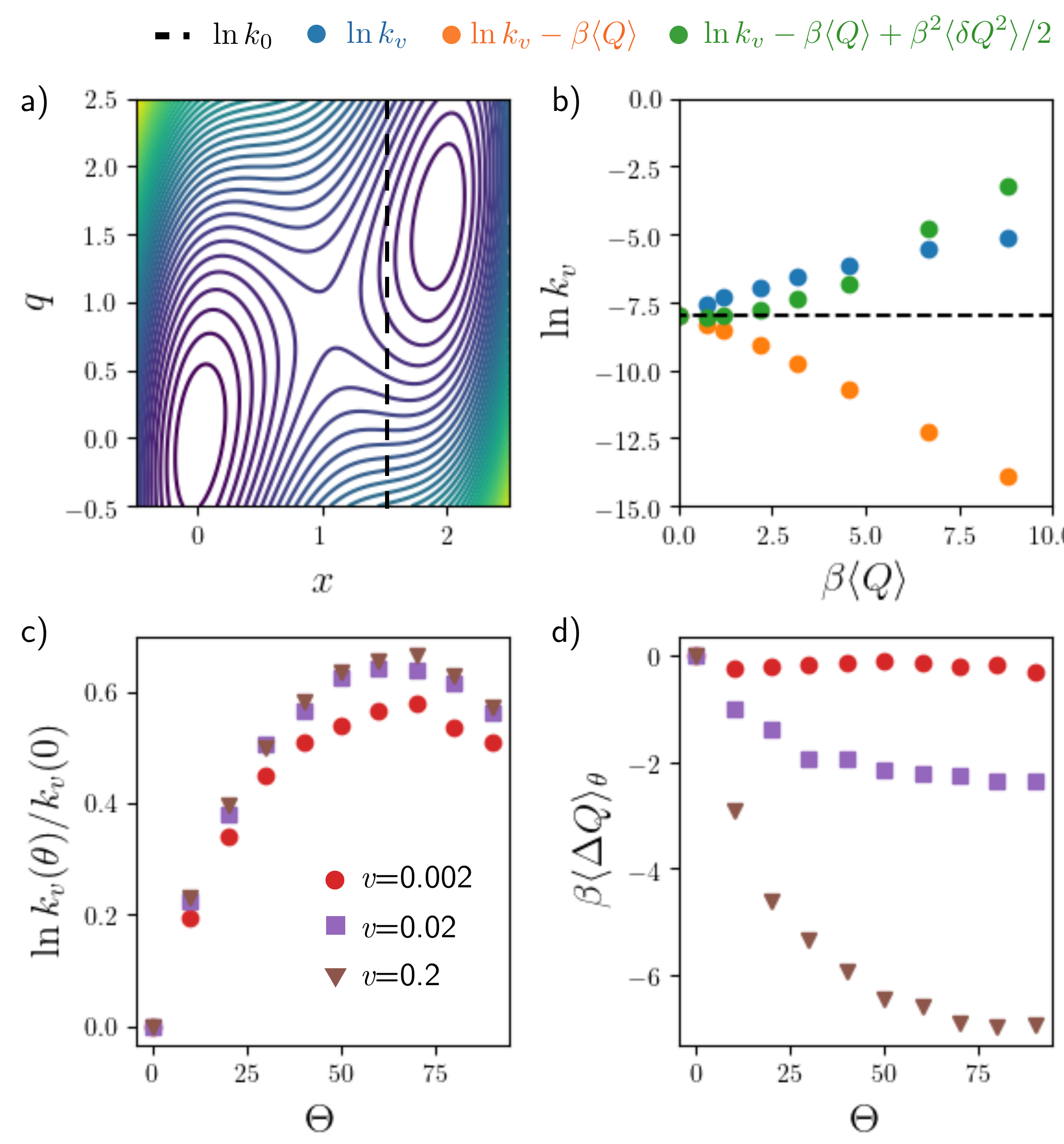}
\caption{Protocol dependence. (a) The potential energy surface in the molecular $x$ and observed $q$ coordinates with absorbing boundary placed at $x^\dag=1.5$. (b) Nonequilibrium rate estimates labeled as in Fig. 1c. (c) Modulation of the nonequilibrium rate with pulling velocity and angle relative to the $q$ axis where the reference rate is taken as pulling along the $q$ direction. (d) Modulation of the dissipated heat with pulling velocity and angle relative to the $q$ axis labeled as in (c) and reference analogously.}
\label{fig2}
\end{figure}

We now consider a simple model often adopted in force spectroscopy studies to understand the role of flexible linkers. Specifically, we consider overdamped motion in 2 spatial dimensions, 
\begin{equation}
\dot{\mathbf{r}} =-\boldsymbol{\mu} \nabla U(\mathbf{r}) + \boldsymbol{\mu} \lambda  + \sqrt{2 \mathbf{D}} \boldsymbol{\eta}
\end{equation}
where $\mathbf{r} =\{q,x \}$, and $q$ is envisioned as a measured extension that is coupled to the true molecular extension, $x$, through a potential
\begin{equation}
U(\mathbf{r}) = \Delta U x^2 (x-2)^2 + \kappa_l (x-q)^2/2 + \kappa_m \mathbf{r}^2 /2
\end{equation}
where  $\Delta U$ denotes the height of the barrier in the molecular coordinate, while $\kappa_l$ and $\kappa_m$ are stiffnesses associated with the linker and trap, respectively.\cite{hinczewski2013mechanical, cossio2015artifacts, berezhkovskii2014multidimensional, manuel2015reconstructing, neupane2016quantifying} In this simplest multi-dimensional model of a single molecule pulling experiment, the molecule undergoes diffusion in the 2-d landscape as shown in Fig. \ref{fig2}a. We perform force-ramp simulations with an added force $\mathbf{\lambda}= a \kappa_m \mathbf{e}_\Theta v t $, parameterized by a pulling vector $\mathbf{e}_\theta=\{ \cos(\Theta),\sin(\Theta )\}$ determined by the angle $\Theta$ relative to the $q$ axis. We fix $ \kappa_m = \kappa_l =  \beta  = \mu_q  = 1$, $x^\dag=1.5$, $\kappa_l=\kappa_m=5$, $\beta \Delta U=5$ and vary $v=\{0,..,0.5 \}$ and $\Theta$.  As before we estimate averages from 10$^4$ trajectories with an absorbing boundary condition at $x^\dag$.

We first consider the inference of the equilibrium rate in the difficult case of pulling along $\Theta=0$ and where the $x$ direction is slow, $\mu_q/\mu_x=20$. Under these conditions, shown in Fig. \ref{fig2}b, we observe that similar to the harmonic oscillator, the dissipative second-cumulant estimate of Eq. \ref{Qexpansion} converges faster to the exact equilibrium rate $\ln k_0$, than either the bare driven rate or the generalized Bell's law estimate, $\ln k_0 \approx \ln k_v - \beta \langle Q\rangle$. As before, the estimate from the mean dissipation converges from below, however in this case the inclusion of the second cumulant results in convergence from above. This is a consequence of the nonlinear system considered in this example, where the non-Gaussian heat statistics manifest an enhanced variance. 

In order to understand the protocol dependence of our rate inference, we consider $\mu_q/\mu_x=1$ and vary $\Theta$ and $v$ in Figs. 2c and 2d. As expected, pulling along the molecular coordinate $x$,  $\Theta=90^\mathrm{o}$, results in a faster process relative to $\Theta=0^\mathrm{o}$, and one which dissipates less heat as quantified by $ \Delta \langle Q \rangle_\Theta =   \langle Q \rangle_\Theta - \langle Q \rangle_0$. The rate within the driven dynamics is maximized near $\Theta=60^\mathrm{o}$, which is in reasonable agreement with the optimal transition state predicted by multidimensional transition state theory.\cite{berezhkovskii2005one} For small $v$, near equilibrium, the dissipated heat approaches zero independent of $\Theta$ as expected for a quasi-reversible process. These results suggest that pulling along any direction may be sufficient to estimate $\ln k_0$, but geometries that minimize the heat until rupture may lead to faster convergence of Eq. \ref{Qexpansion}.

\begin{figure}
\centering
\includegraphics[width=8.5cm]{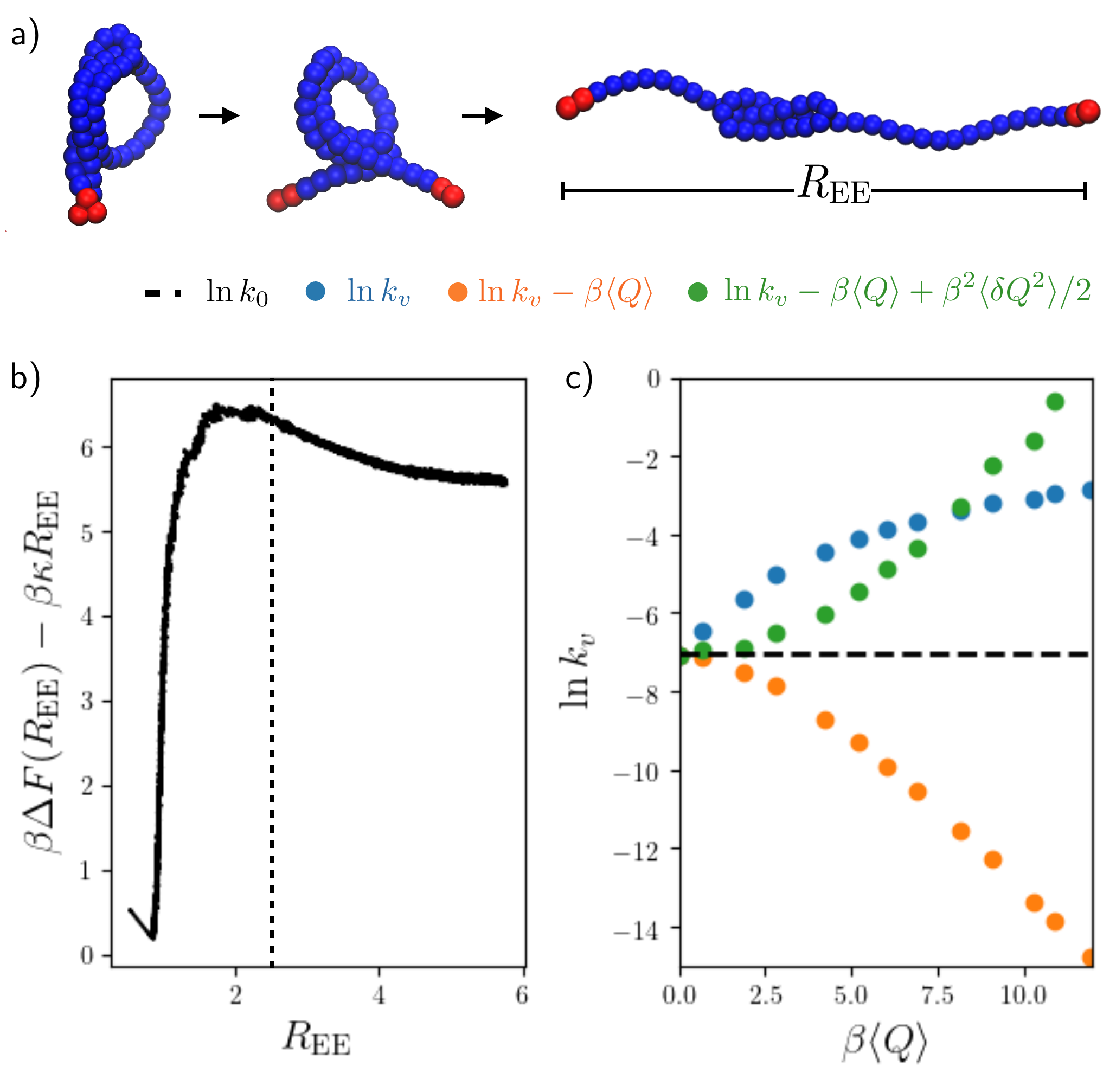}
\caption{ (a) A semi-flexible polymer with reactive ends, highlighted in red, mechanically unfolded to a large end to end distance during a force ramp experiment. (b) The free energy of the untethered polymer as a function of the end to end distance. The black dashed line denotes the absorbing boundary condition. (c) Nonequilibrium rate estimates, labeled as in Fig 1c, as a function of heat.}
\label{fig3}
\end{figure}

As an example of a non-linear many-particle system, we consider stretching a semi-flexible polymer with reactive ends\cite{wu2018role} that attract each other with a strong, short-ranged potential. 
The configuration of the polymer consists of 3$N$ dimensions, $\mathbf{r}=\{\mathbf{r}_1, \dots, \mathbf{r}_N\}$, and evolves with underdamped Langevin dynamics  
\begin{equation}
     m \mathbf{\ddot{r}}_i =  - \mu^{-1}  \mathbf{\dot{r}}_i  - \boldsymbol{\nabla}_i U(\mathbf{r})  +  \boldsymbol{\lambda} + \mu^{-1} \sqrt{2 D} \boldsymbol{\eta_i}. 
\end{equation}
where $m$ is the mass of a monomer. The monomers interact through a potential that consists of $U(\mathbf{r})=U_\mathrm{b}(\mathbf{r})+U_\mathrm{a}(\mathbf{r})+U_\mathrm{nb}(\mathbf{r})$ where $U_\mathrm{b}(\mathbf{r})$ is a harmonic bond potential between adjacent monomers
\begin{equation}
     U_\mathrm{b}(\mathbf{r}) = \frac{\kappa_\mathrm{b}}{2} \sum_{i=1}^{N-1} \left ( \mathbf{r}_{i+1}-\mathbf{r}_i \right )^2
\end{equation}
with stiffness $\kappa_\mathrm{b}$ and $U_{\mathrm{a}}(\mathbf{r})$ is a harmonic angular potential that penalizes bending
\begin{equation}
     U_{\mathrm{a}}(\mathbf{r}) = \frac{\kappa_\mathrm{a}}{2}\sum_{i=2}^{N-1} \left ( \frac{\mathbf{r}_{i+1,i} \cdot \mathbf{r}_{i,i-1} }{{r}_{i+1,i} {r}_{i,i-1} } -1 \right )^2
\end{equation}
where $\mathbf{r}_{i,i-1}=\mathbf{r}_{i}-\mathbf{r}_{i-1}$ is the vector between two adjacent monomers, ${r}_{i,i-1}$ denotes its magnitude, and the potential has a stiffness $\kappa_\mathrm{a}$.  The nonbonding potential $U_\mathrm{nb}(\mathbf{r})$ has a short ranged form,
\begin{equation}
     U_{\mathrm{nb}}(\mathbf{r}) = \frac{1}{2} \sum_{i \ne j=1}^N  \epsilon_{ij} \left[5 \left(\frac{\sigma}{r_{i,j}}\right)^{12}- 6 \left(\frac{\sigma}{r_{i,j}}\right)^{10}\right]
\end{equation}
where $\sigma$ sets the characteristic size of a monomer and $\epsilon_{ij}$ the interaction strength between monomer $i$ and $j$. 
In order to model a reactive linker, we set the two monomers on each end to interact more strongly than the monomers in the interior of the polymer, and the cross interactions are neglected. We hold the 1st monomer fixed at the origin, so the end to end vector is $\mathbf{R_{EE}} = \mathbf{r_{N}}$, and pull the two ends apart with $\boldsymbol{\lambda}= -\kappa (\mathbf{R_{EE}}- v t \mathbf{\hat{x}}) $. Characteristic snapshots are shown in Fig.~\ref{fig3}a). We adopt a unit system with $\beta=m=\sigma=1$, so that $\tau=1/\sqrt{\beta m \sigma^2}$. We set $\mu=5$, $\kappa_\mathrm{b}=110$, $\kappa_\mathrm{a}=4.5$, $\kappa=5.5$, end monomer interactions $\epsilon_{1,N}=4.5$, and interior monomer interactions $\epsilon_{3,N-3}=1.7$, $N=50$. We employ a time step of 10$^{-2}\tau$ and estimate averages from 500 trajectories. 

Under these conditions, the semi-flexible polymer permits two types of conformations--  a folded state for small $\mathbf{R_{EE}}$ where the linker monomers are bound, and an unfolded state for large $\mathbf{R_{EE}}$ where the linker monomers do not interact strongly. To differentiate between these two regions, we first computed the work to pull $\mathbf{R_{EE}}$ reversibly, denoted by $\beta \Delta F(\mathbf{R_{EE}})$. This is shown in Fig.~\ref{fig3}b), and evaluated using the Jarzynski equality\cite{jarzynski1997nonequilibrium} within a steered molecular dynamics framework.\cite{park2004calculating} For the free energy calculation we employed the same constant $v$ protocols, including all of the simulation data shown in Fig.~\ref{fig3}c). The free energy exhibits a deep minima for small end to end distances and a second shallow basin for large end to end distances. 
Under a small additional load, $-\kappa _\mathrm{EE} \mathbf{R_{EE}}$, the two basins are separated by a barrier in the free energy. Using this biased free energy we set an absorbing boundary condition for our pulling calculation to $\mathbf{R_{EE}}^\dag=2.5 \sigma$.

We pulled the polymer at loading rates over the interval $v = \{0.001-0.2\} \sigma/\tau$ and measured the dissipative heat and first passage time to $\mathbf{R_{EE}}^\dag$. Shown in Fig.~\ref{fig3}c), we find that convergence to the true equilibrium rate is fastest using Eq.~\ref{Qexpansion}, and as in the 2d model the estimate converges from above. As in both previous models, the incorporation of the variance of the heat provides an accurate estimate of the equilibrium over a range of heats that is comparable to a significant fraction of the native barrier, in this case as long as $\beta \langle Q \rangle < 3$. The first order estimate converges to the true equilibrium rate slowly from below. Over the range of pulling velocities considered, if either the Bell's law estimate or the bare rate were fit and linearly extrapolated to $v=0$, both would provide an estimate of the equilibrium rate that is about an order of magnitude too high. 

Taken together, our results demonstrate an underlying stochastic thermodynamic basis for Bell's law under nonequilibrium driving and a useful means for going beyond it to infer equilibrium transition rates. Within the context of single molecule force ramp experiments, we have demonstrated a robust way to infer unfolding rates using the first two cumulants of the heat distribution, conditioned on ending at an absorbing transition state. In the future, this perturbative response method may be used to study the rare kinetics of more detailed protein models, protein unfolding in optical tweezing and atomic force microscopy experiments, and other molecular transitions that can be sped up by applied force.\cite{bustamante2021optical, petrosyan2021single,wang2021theory,zhang2015statistical} The nonequilibrium thermodynamic framework developed here works not only with constant velocity force ramps, but could be used with more complex protocols. Indeed, protocols can be optimized to allow for rate inferences,\cite{das2022direct} or could be used as a theoretical framework for understanding other approximate methods that use driven dynamics to infer rates with applied force.\cite{voter1997hyperdynamics,pena2022assessing,khan2020fluxional,tiwary2013metadynamics}         
\\

{\bf Acknowledgements} The authors would like to thank Glen Hocky and Carlos Bustamante for helpful discussions during the preparation of this manuscript. This work has been supported by NSF Grant CHE-1954580

\section*{References}
\bibliography{rate_amp_diss}

\end{document}